\documentclass[conference]{IEEEtran}
\IEEEoverridecommandlockouts
\usepackage{cite}
\usepackage{amsmath,amssymb,amsfonts}
\usepackage{algorithmic}
\usepackage{graphicx}
\usepackage{textcomp}
\usepackage{xcolor}
\usepackage{float}
\usepackage{makecell} 
\usepackage[utf8]{inputenc}
\usepackage{enumitem}
\def\BibTeX{{\rm B\kern-.05em{\sc i\kern-.025em b}\kern-.08em
    T\kern-.1667em\lower.7ex\hbox{E}\kern-.125emX}}
\begin{document}

\title{Machine Learning-Based Anomaly Detection of Correlated Sensor Data: An Integrated Principal Component Analysis-Autoencoder Approach\\
{\footnotesize \textsuperscript{*}}
}

\makeatletter
\newcommand{\linebreakand}{%
  \end{@IEEEauthorhalign}
  \hfill\mbox{}\par
  \mbox{}\hfill\begin{@IEEEauthorhalign}
}
\makeatother

\author{\IEEEauthorblockN{Tanish Baranwal}
\IEEEauthorblockA{\textit{DomaniSystems, Inc.} \\
Shelton, CT 06484 \\
tekotan@berkeley.edu}
\and
\IEEEauthorblockN{Arnab Das}
\IEEEauthorblockA{\textit{Johns Hopkins University Applied Physics Laboratory} \\
Laurel, MD, 20723 \\
arnab.das@jhuapl.edu} \\
\and
\IEEEauthorblockN{Srihari Varada}
\IEEEauthorblockA{\textit{DomaniSystems, Inc.} \\
Shelton, CT 06484 \\
varada@ieee.org} \\
\and
\IEEEauthorblockN{Santanu Das}
\IEEEauthorblockA{\textit{DomaniSystems, Inc.} \\
Shelton, CT 06484 \\
das@domanisystems.com}
\and 
\IEEEauthorblockN{Mohammad R. Haider}
\IEEEauthorblockA{\textit{University of Missouri} \\
\textit{Department of Electrical Engineering and Computer Science}\\
Columbia, MO 65211 \\
mhaider@missouri.edu}
}

\maketitle

\begin{abstract}
  The growing adoption of IoT systems in industries like transportation, banking, healthcare, and smart energy has increased reliance on sensor networks. However, anomalies in sensor readings can undermine system reliability, making real-time anomaly detection essential. While a large body of research addresses anomaly detection in IoT networks, few studies focus on correlated sensor data streams, such as temperature and pressure within a shared space, especially in resource-constrained environments. To address this, we propose a novel hybrid machine learning approach combining Principal Component Analysis (PCA) and Autoencoders. In this method, PCA continuously monitors sensor data and triggers the Autoencoder when significant variations are detected. This hybrid approach, validated with real-world and simulated data, shows faster response times and fewer false positives. The F1 score of the hybrid method is comparable to Autoencoder, with much faster response time which is driven by PCA.
\end{abstract}


\section{Introduction}
Internet of Things (IoT)-based systems are increasingly utilized across applications such as transportation, banking, healthcare, and smart energy \cite{item_1,item_2}, relying on sensor networks to collect and transmit data to a central computing resource for seamless operations and data-driven insights. However, anomalies in sensor data—caused by environmental factors, hardware malfunctions, or cyber threats—can undermine the reliability and security of these systems. Detecting and mitigating such anomalies in real-time is critical for ensuring system integrity and operational effectiveness.

Existing research on IoT anomaly detection mainly focuses on network-level anomalies, such as abnormal traffic, latency, and cyber threats \cite{item_3,item_4,item_5}. However, there is limited research on detecting anomalies in correlated sensor data streams, particularly in resource-constrained environments. Many IoT systems use multiple sensors to measure the same or related parameters, such as temperature and pressure. For example, modern aircraft use multiple altimeters (barometric and radar) to cross-check altitude during critical phases like landing. Current research lacks effective methods for detecting anomalies in such correlated sensor data streams, particularly when sensors are spatially separated or collect data at varying speeds or intervals, as seen in electric vehicles or industrial machines with limited processing power and memory.

Traditional statistical methods \cite{item_5,item_6}, like time-series analysis, are computationally intensive and require large historical datasets, making them unsuitable for real-time detection in resource-constrained environments. Machine learning (ML) and deep learning (DL) approaches \cite{item_7,item_8,item_9,item_10,item_12} have shown promise in anomaly detection but often require substantial computational resources and labeled data, which may not always be available. Additionally, many ML models struggle to differentiate between true anomalies and benign variations, leading to higher false positive rates. This highlights the need for an approach that balances detection speed with computational efficiency, especially in resource-constrained IoT systems.

To address these challenges, this paper proposes a novel hybrid machine learning approach combining PCA \cite{item_13} and Autoencoders \cite{item_14} for anomaly detection in correlated sensor data streams. PCA, a linear dimensionality reduction technique, continuously monitors sensor data for deviations with minimal computational overhead. When significant variations are detected, an Autoencoder—a deep learning model that captures nonlinear dependencies—further analyzes the data to confirm anomalies. This adaptive approach ensures real-time detection, minimizes false positives, and optimizes resource usage.

The effectiveness of the proposed method is validated with both real-world and synthesized sensor data. Simulation results show that the hybrid approach reduces computational load, improves response time, and lowers false positive rates compared to standalone PCA or Autoencoder models.

The paper is structured as follows: Section \ref{sec:motivation} discusses the motivation for our approach, Section \ref{sec:methods} presents the methodology and experimental setup, Section \ref{sec:results} provides quantitative analyses and results, and Section \ref{sec:conclusion} concludes with key findings.

\section{Motivation for Technical Approach} \label{sec:motivation}

The motivation behind this research is to develop a machine learning-based approach for IoT anomaly detection, addressing the challenge of high-dimensional, correlated sensor data. Among dimensionality reduction techniques, PCA, UMAP \cite{item_15}, TSNE \cite{item_16}, and Autoencoders are very popular and widely used. A comparative overview is provided in Table \ref{fig:table1}, where `n' refers to the training dataset size, `input' is the window size, and `output' denotes the latent space dimension.

\begin{table}[!t]
  \caption{Comparison of Dimensionality Reduction Techniques}
  \label{fig:table1}
  \centering
  \begin{tabular}{l l l l}
  \hline
  \textbf{Technique} & \textbf{Purpose} & \makecell[l]{\textbf{Computational} \\ \textbf{Complexity}} & \makecell[l]{\textbf{Retraining}\\\textbf{Needed}} \\
  \hline
  t-SNE &
  \makecell[l]{2D/3D\\Visualization} &
  O(\(n^2\)) &
  Yes \\
  
  UMAP &
  Clustering &
  O(\(n^2\)) &
  Yes \\
  
  PCA &
  \makecell[l]{Linear\\Compression} &
  O(\(\text{input} \times \text{output}\)) &
  No \\
  
  Autoencoder &
  \makecell[l]{Nonlinear\\Compression} &
  Constant &
  No \\
  \hline
  \end{tabular}
  \end{table}

This study focuses on PCA and Autoencoders, which are particularly useful for feature extraction in high-dimensional spaces \cite{item_17,item_18}, while TSNE and UMAP are mainly used for 2D/3D visualization. TSNE and UMAP however require retraining for new data points, making them less efficient for real-time updates. Thus, the comparative analysis in Section IV compares PCA, Autoencoder, and the proposed hybrid model against UMAP.

The Autoencoder used in this research is based on a Long Short-Term Memory (LSTM) \cite{item_18} encoder-decoder architecture, ideal for processing temporal data and capturing long-term dependencies. To evaluate these techniques, we apply them to a sliding window of 100 historical time samples, reducing the dataset’s dimensionality. We then compute a Euclidean distance matrix, containing the pairwise distances between sensors in matrix form, to quantify pairwise relationships between encoded sensor data. Significant deviations in these distances indicate anomalies. A faulty sensor will show abnormal distances with all other sensors, while functioning sensors exhibit only minor deviations.

This approach ensures robust anomaly detection by leveraging both linear and nonlinear dimensionality reduction techniques.

\section{Methods}\label{sec:methods}

The Methods section is structured into three distinct subsections, each corresponding to the respective sections in the RESULTS. First, we outline the methodology for assessing the effectiveness of PCA and Autoencoders in anomaly detection. Second, we present the approach for the quantitative analysis of PCA and Autoencoder performance. Finally, we introduce the proposed hybrid model and describe the evaluation framework employed to assess its effectiveness.

\subsection{Methods for characterizing the effectiveness of PCA vs. Autoencoder in Anomaly Detection} \label{sec:methods1}

We used sensor data from Intel Berkeley Research Labs \cite{item_19}, collected from 54 sensors deployed between February 28th and April 5th, 2004. Mica2Dot sensors \cite{item_20} with weatherboards recorded humidity, temperature, light, and voltage every 31 seconds using the TinyDB \cite{item_21} in-network query system on the TinyOS platform \cite{item_22}. Data from 4 temperature sensors in the same lab area was selected, with readings taken over 2000 time steps. At the 1000th time step, Sensor 4 was deliberately triggered to manifest anomalous behavior.

Two anomaly models were studied: (a) an anomaly resulting from a sudden shift in the mean of readings of a sensor in a group of n sensors, where n=4 in this experiment (b) an anomaly resulting from one of the sensors in a group exhibiting `erasure error' (certain readings being set to zero).

In anomaly scenario (a), sensor four’s readings were shifted after the 1000th time step by adding a scalar value to its last 1000 readings, altering the mean, while keeping the standard deviation and distribution unchanged. The other three sensors remained normal over 2000 time steps. To assess the sensitivity of PCA and Autoencoder methods, two experiments were conducted: one with a deviation of 5 (adding 5 to the anomalous readings) and another with a deviation of 40 (adding 40 to the anomalous readings).

In anomaly scenario (b), the sensor four was simulated to exhibit erasure error (i.e., certain readings being zeroed out). To compare the sensitivity of the PCA and Autoencoder approaches similarly to the first model, two trials were simulated: in one trial, a 5\% erasure error was applied (which means on average one in 20 readings get zeroed out), and in the other, a 20\% erasure error was applied.

\subsection{Methods for quantitative analysis of Autoencoder and PCA}\label{sec:methods2}

We also use numerical metrics such as F1 score, precision, and recall \cite{item_23} to compare their effectiveness. The models are also tested across various distributions, probabilities of anomalous behavior, and mean deviations. The quantitative analysis includes: 1) sensitivity analysis of PCA and Autoencoder using simulated data, 2) comparison of a hybrid model with other dimensionality reduction approaches using Intel Berkeley Research Labs data and 3 runtime analysis of the hybrid model with the same dataset.

A sensitivity analysis was conducted using simulated Gaussian data to enhance experimental control. To ensure the model's performance was representative of real-world conditions, the same analysis was applied to both real-world and simulated datasets. Specifically, Gaussian distributions with fixed means and standard deviations were sampled for all four sensors. The model demonstrated comparable performance across both datasets when evaluated using identical anomaly detection models.

Anomaly detection was performed by computing Euclidean distances between sensor pairs and applying a binary classification (0: no anomaly, 1: anomaly), flagging sensors as anomalous if distances exceeded a set threshold. For the Autoencoder, anomalies were additionally detected based on elevated reconstruction loss relative to training data distribution, using a loss threshold approach consistent with established methodologies \cite{item_24}.

Precision, recall, and F1 scores were calculated to assess PCA and Autoencoder anomaly detection sensitivity and accuracy. Four types of anomalies (Poisson, Normal, Uniform, and single-point) were evaluated using a fixed standard deviation of 5, across odd mean values from 1 to 100. Metric values near 1 indicated anomalies, while values near 0 indicated normal behavior.

The experimental setup included four sensors over 500 time steps. Three sensors functioned normally throughout, sampling from a Gaussian distribution (mean=50, SD=5), while the fourth sensor introduced anomalies at step 250, as depicted in Figure \ref{fig:simflowchart}.

\begin{figure}[h]
  \centering
  \includegraphics[scale=0.3]{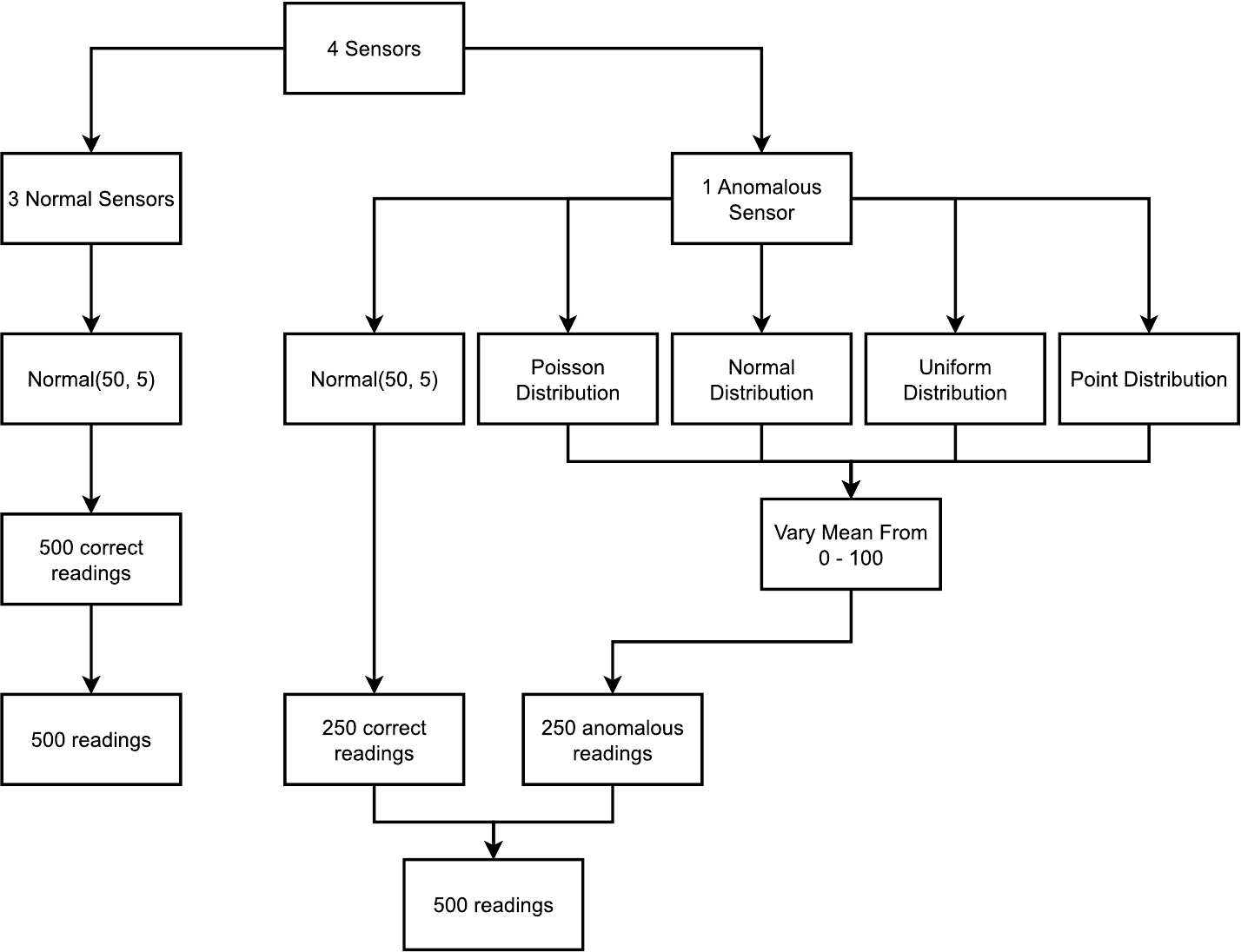}
  \caption{Setup for simulation of normal and anomalous behavior}
  \label{fig:simflowchart}
  \end{figure}

\subsection{Proposed Hybrid Model and its Evaluation}\label{sec:methods3}
The primary motivation for employing the Hybrid model is to leverage the advantages of the Autoencoder while improving response time. As illustrated in Figure \ref{fig:hybridflowchart}, the hybrid approach integrates the computational efficiency of PCA with the robust anomaly detection capabilities of the Autoencoder, thereby improving overall performance. In this framework, PCA serves as an initial filtering mechanism to identify potential anomalies, which are subsequently verified by the Autoencoder at each time step flagged by PCA. Given that PCA is susceptible to false positives, the Autoencoder is invoked more frequently when anomalies are detected. The average response time of the Hybrid model is dependent upon the frequency with which the Autoencoder is activated.

\begin{figure}[h]
  \centering
  \includegraphics[scale=0.4]{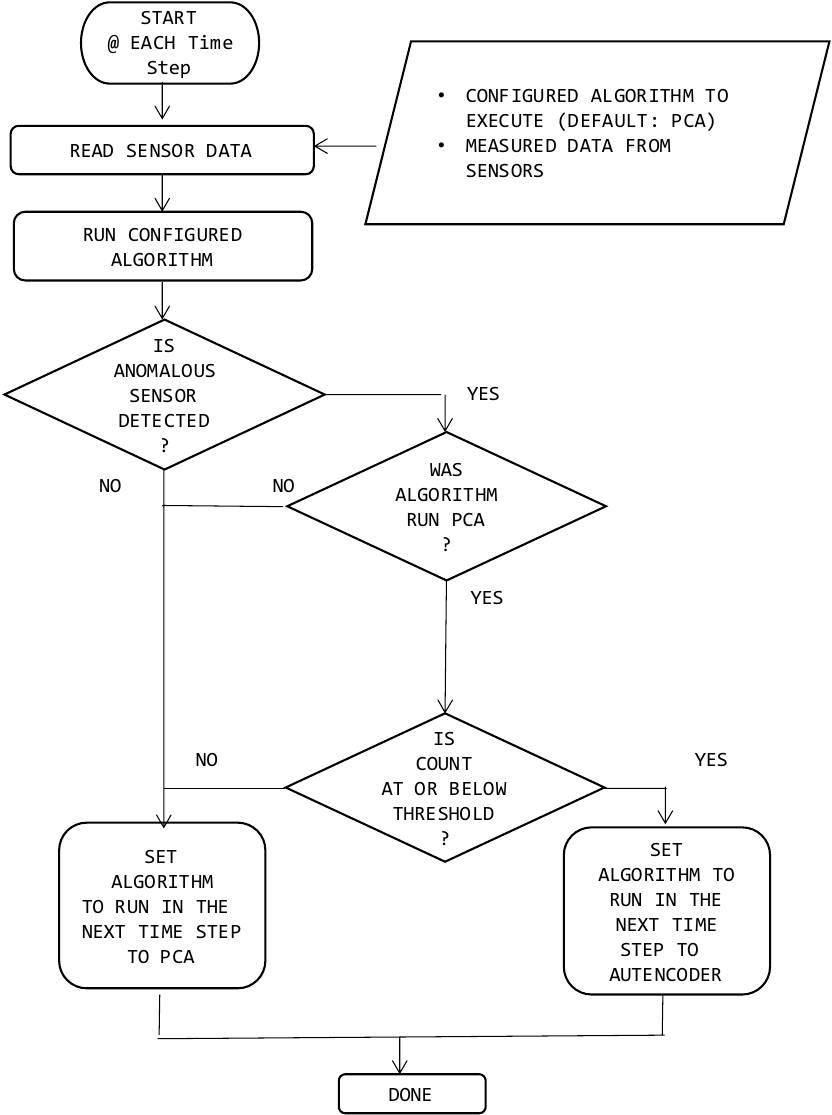}
  \caption{Hybrid method for the detection of anomalous sensor}
  \label{fig:hybridflowchart}
  \end{figure}

We assess the effectiveness of the Hybrid model by comparing its performance against PCA, Autoencoder, and UMAP, using the F1 score as a benchmark for both accuracy and computational efficiency. TSNE is excluded due to its inefficiency, as it requires retraining at each iteration. The evaluation is conducted on 6000 time steps of sensor data from the Intel Berkeley Research Labs focusing on four highly correlated sensors. To introduce controlled anomalies in the fourth sensor, its readings are segmented into bins of 1000, with the mean artificially shifted by 10 in the odd-numbered bins, rendering 50\% of the data anomalous. Each model is rigorously evaluated based on its F1 score, while the runtime per time step is systematically measured for comparison among the models.

The reduction of the response time is quantified when using the Hybrid model by comparing the average response time of the Autoencoder model versus the Hybrid model at different rates of anomaly in the data. Using the same 6000 time steps of the Intel Berkeley Research Labs sensor data, 10 trials were simulated, with the trials ranging from having 20\% to 100\% anomalies in the fourth sensor. The average response time for the Autoencoder model and the Hybrid model are computed on each trial. The run times were determined on an AMD Ryzen 9 CPU.

\section{Results}\label{sec:results}

We break up the results section into three sections, corresponding to the three sections in the methods section. We first discuss (a) the effectiveness of PCA vs. Autoencoder in Anomaly Detection, (b) a quantitative analysis of Autoencoder and PCA with regards to different distributions and mean differences, and (c) the evaluation of the proposed hybrid model.

\subsection{Effectiveness of PCA vs. Autoencoder in Anomaly Detection}\label{sec:results1}

As discussed in the methods, we compare PCA and Autoencoder performance on two anomaly models: anomaly model (a) corresponds to a shift in mean, and anomaly model (b) corresponds to random erasure error.

\subsubsection{Anomaly Model (a) Corresponding to a Shift in Mean}\label{sec:results1a}

Anomaly scenario (a) assumes the fourth sensor experiences a sudden malfunction causing a positive shift in readings, changing the mean by either 5 or 40. These values were chosen to analyze effects of small versus large deviations.

Figures \ref{fig:autoencoder-a} and \ref{fig:pca-a} illustrate distance matrices for the Autoencoder and PCA, respectively. Highlighted areas depict increased distances at the midpoint (1000th timestep), corresponding to shifts in mean between the anomalous sensor (sensor 4) and the normal sensors.

\begin{figure}[h]
  \centering
  \includegraphics[scale=0.26]{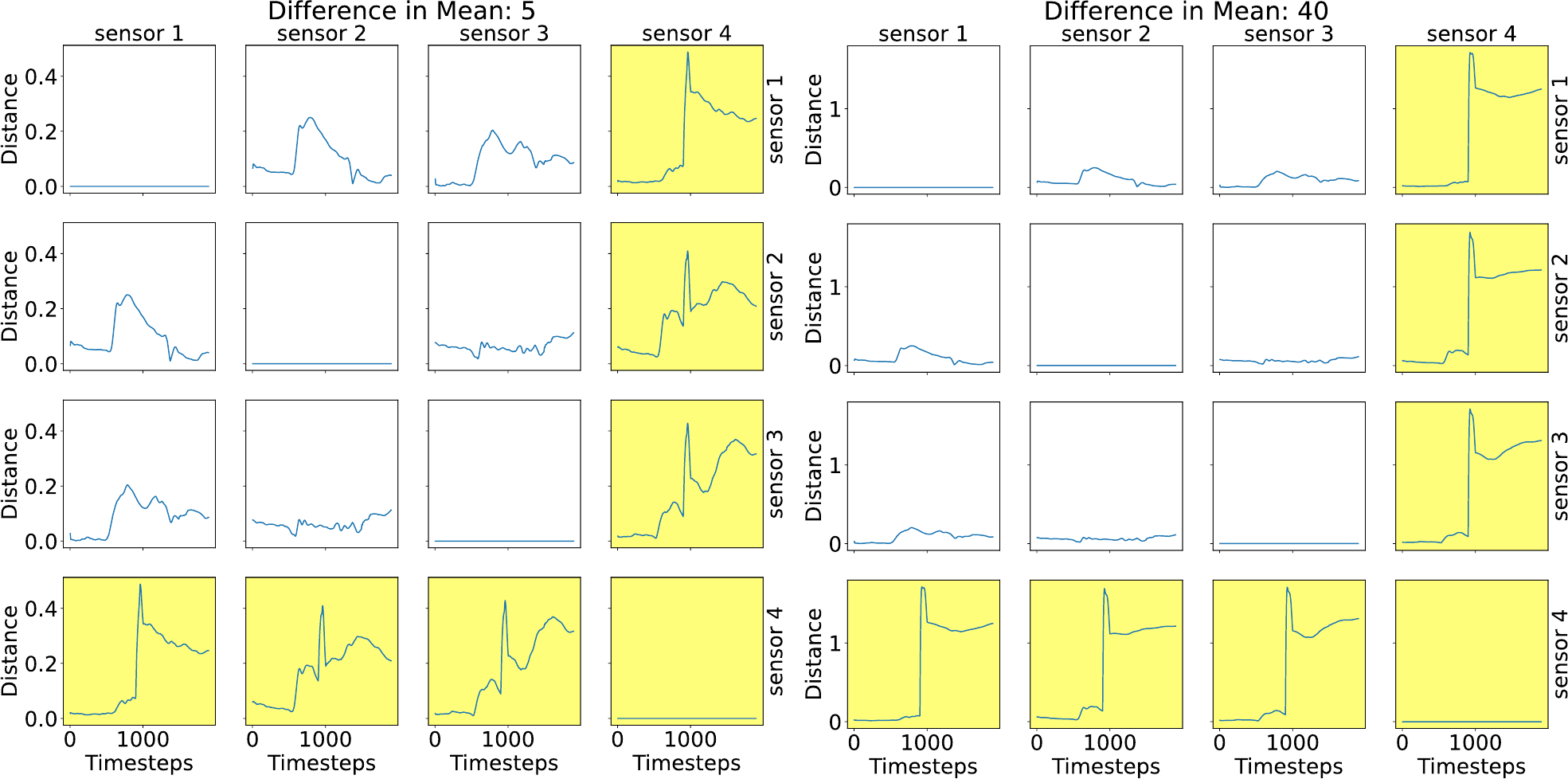}
  \caption{Autoencoder simulation results for anomaly model (a)}
  \label{fig:autoencoder-a}
  \end{figure}

At timestep 1000, sensor 4's anomalous behavior immediately alters sensor distances. A mean deviation of 5 significantly impacted PCA but had minimal consistent effect on the Autoencoder, indicating PCA's higher sensitivity. A deviation of 40 resulted in significant responses from both methods. PCA's linear nature yields proportionally increasing responses, while the Autoencoder responds only when deviations surpass a learned threshold, demonstrating its robustness to minor variations.

\begin{figure}[h]
\centering
\includegraphics[scale=0.26]{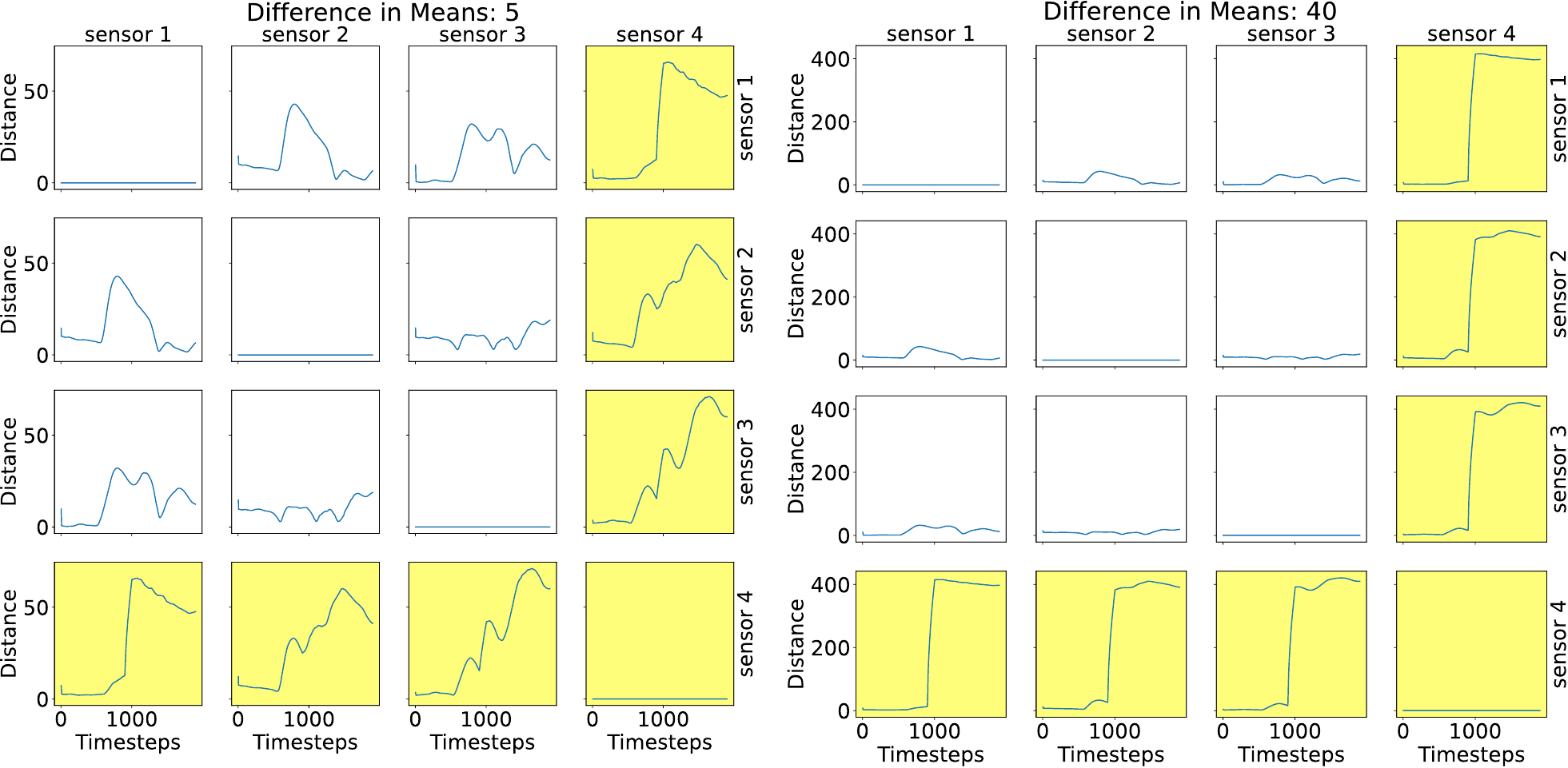}
\caption{PCA simulation results for anomaly model (a)}
\label{fig:pca-a}
\end{figure}

\subsubsection{Anomaly Model (b) Corresponding to Erasure Error}\label{sec:results1b}

In this anomaly scenario, sensor four exhibited erasure errors, with certain readings zeroed out. Two trials evaluated PCA and Autoencoder sensitivity: one with 5\% erasure (one in 20 readings zeroed) and another with 20\% erasure.

Figures \ref{fig:autoencoder-b} and \ref{fig:pca-b} depict distance matrices for Autoencoder and PCA under both erasure conditions. The Autoencoder approach significantly responded to both 5\% and 20\% erasure errors, whereas PCA only showed significant response at 20\%. PCA distances increased linearly with erasure rate, while the Autoencoder showed comparable responses at both rates, indicating its robust learning capability in distinguishing anomalies.

\begin{figure}[h]
  \centering
  \includegraphics[scale=0.26]{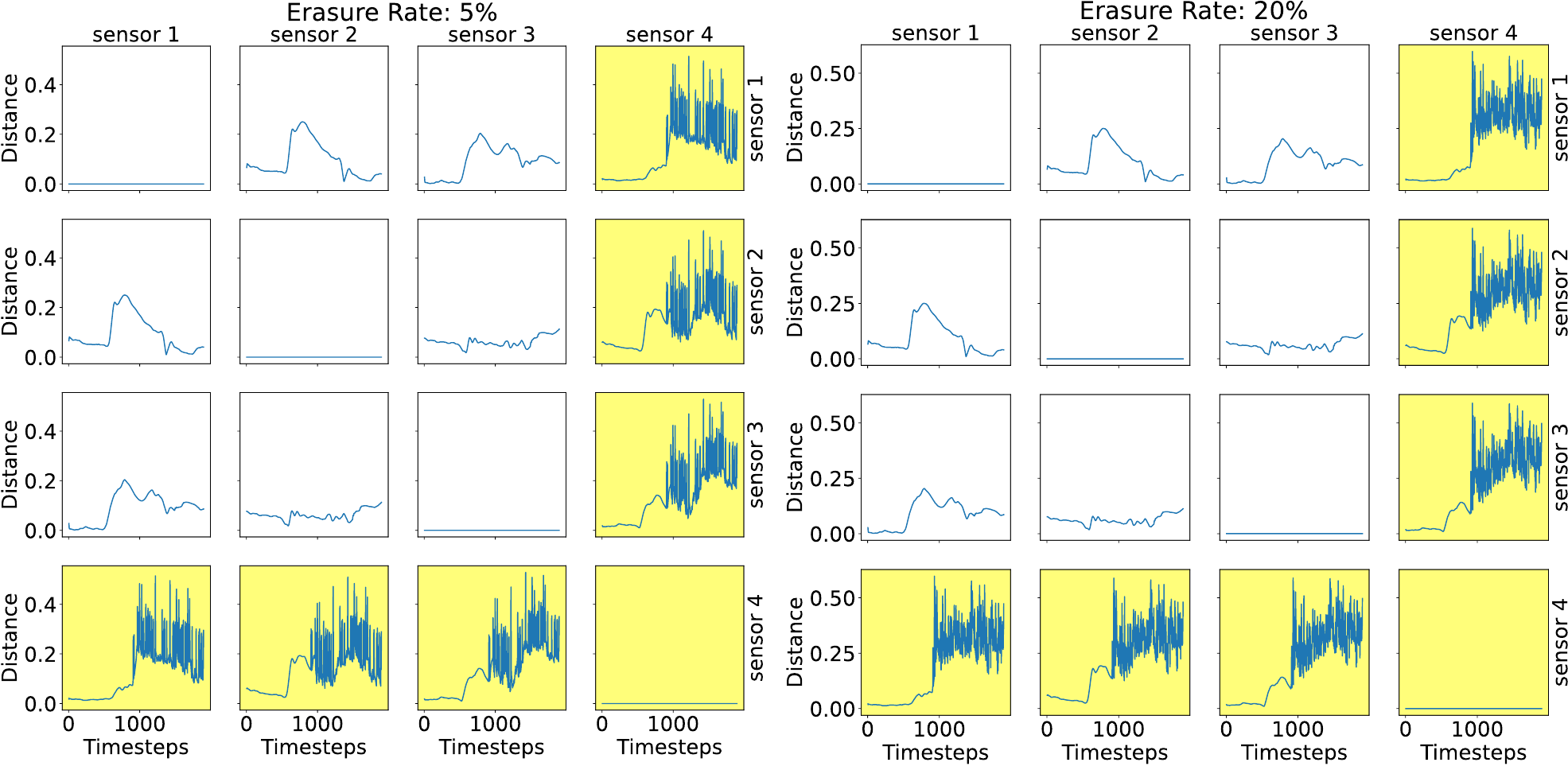}
  \caption{Autoencoder simulation results for anomaly model (b)}
  \label{fig:autoencoder-b}
  \end{figure}

\begin{figure}[h]
  \centering
  \includegraphics[scale=0.26]{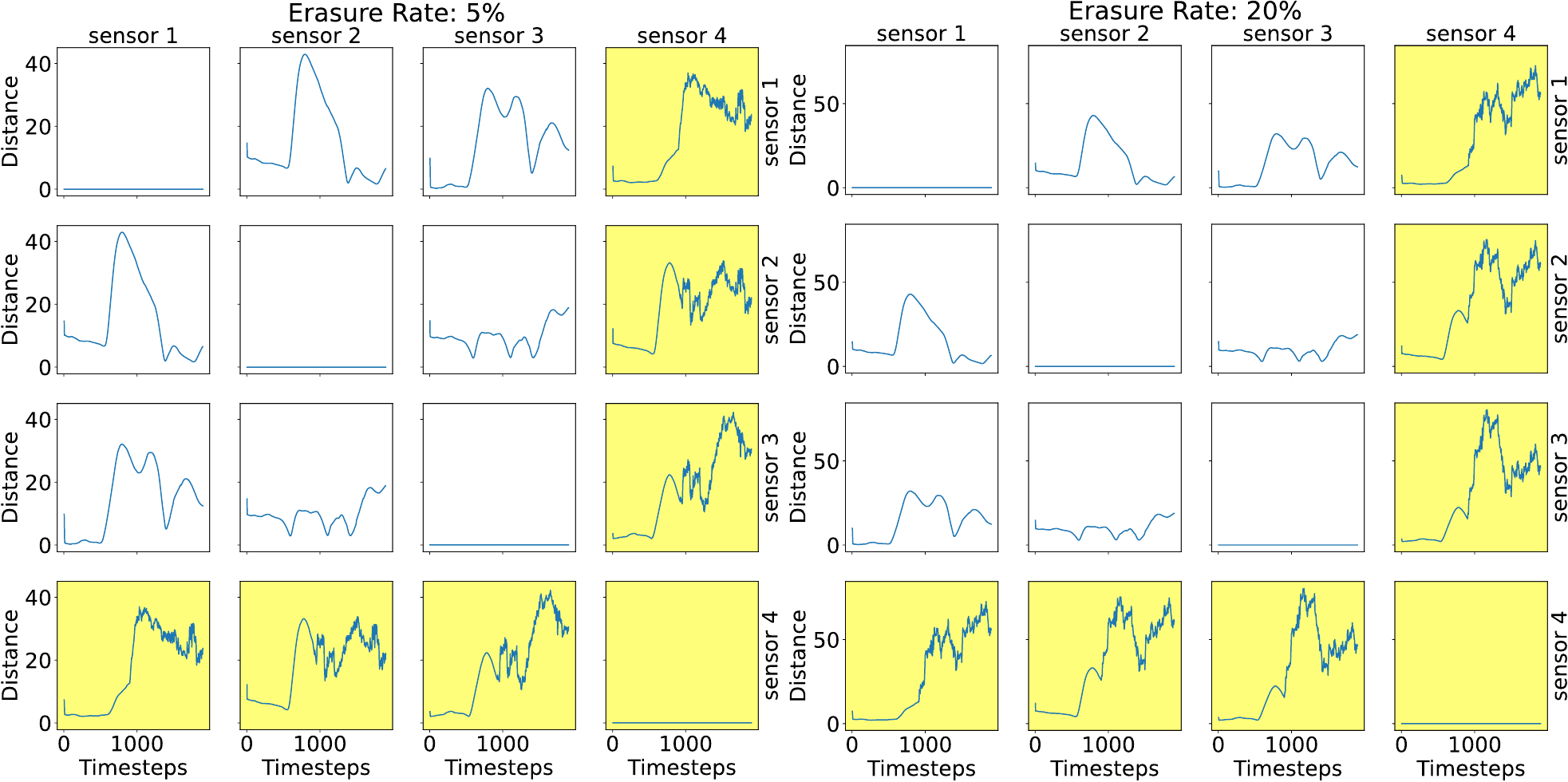}
  \caption{PCA simulation results for anomaly model (b)}
  \label{fig:pca-b}
  \end{figure}

Erasure errors notably increased the standard deviation of Autoencoder-encoded distances but minimally affected PCA distances. Thus, the Autoencoder demonstrated higher sensitivity to erasure errors due to their influence on sensor reading distributions.

\subsection{Quantitative Analysis of Autoencoder and PCA with regards to Different Distributions and Mean Differences}

In Figures \ref{fig:autoencoder-mean-sens} and \ref{fig:pca-mean-sens}, the x-axis represents the mean of the anomalous readings for each trial, and the y-axis shows the precision, recall, and F1 score values.

\begin{figure}[h]
  \centering
  \includegraphics[scale=0.275]{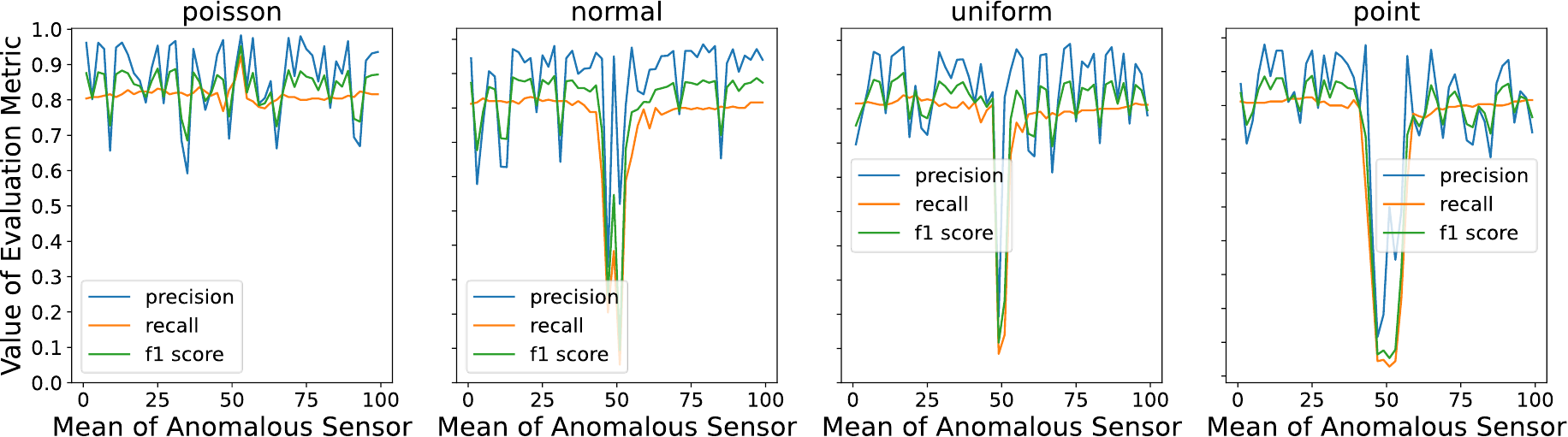}
  \caption{Sensitivity and accuracy of the Autoencoder (for the fourth sensor)}
  \label{fig:autoencoder-mean-sens}
  \end{figure}

\begin{figure}[h]
  \centering
  \includegraphics[scale=0.275]{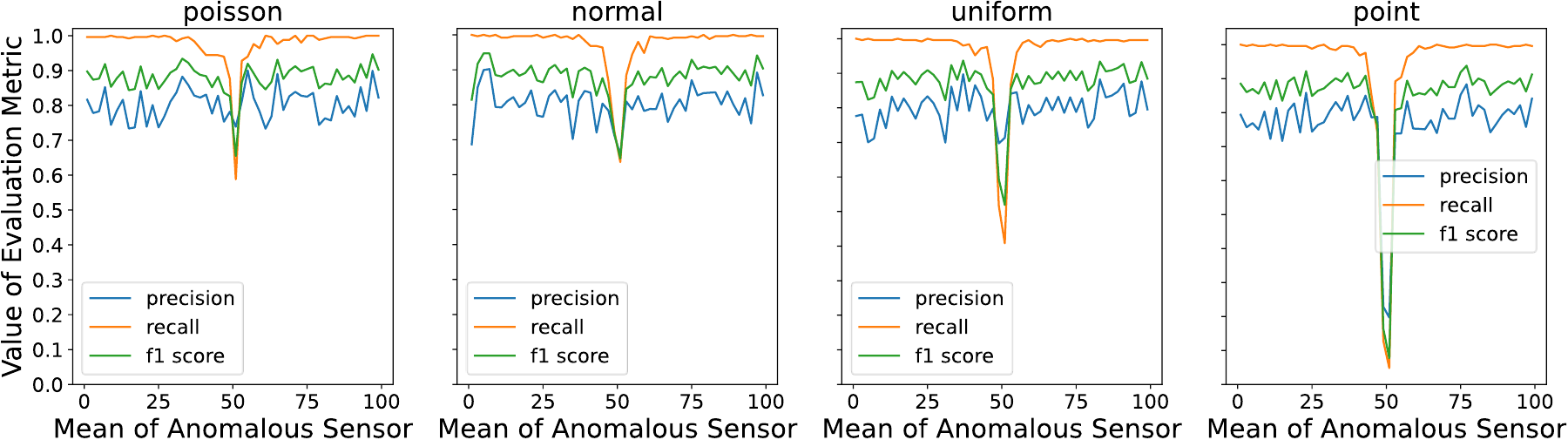}
  \caption{Sensitivity and accuracy of the PCA (for the fourth sensor)}
  \label{fig:pca-mean-sens}
  \end{figure}

For all trials, the second half of the fourth sensor’s readings are considered anomalous (labeled as 1), even when the mean is close to normal values. Evaluation metric values closer to 1 indicate an anomaly, and values closer to 0 suggest no anomaly.

This approach allows analysis of sensitivity and false positive rates within the Range of Practical Equivalence (ROPE)\cite{item_25}, defined as (45, 55), to account for noise or subtle environmental differences between sensors.

In Figure \ref{fig:autoencoder-mean-sens}, the Autoencoder’s lower F1 scores for “means” (averages) within the ROPE for the normal, uniform, and point distributions reflect its lower sensitivity to small mean changes and fewer false positives in the ROPE. This is expected, as readings with a mean near 50 should not be classified as anomalous.

In contrast, Figure \ref{fig:pca-mean-sens} shows that PCA exhibits fewer dips in F1 score across the four distributions, leading to more false positives. PCA continues to classify readings with means close to the true sensor values as anomalous.

\begin{figure}[h]
  \centering
  \includegraphics[scale=0.275]{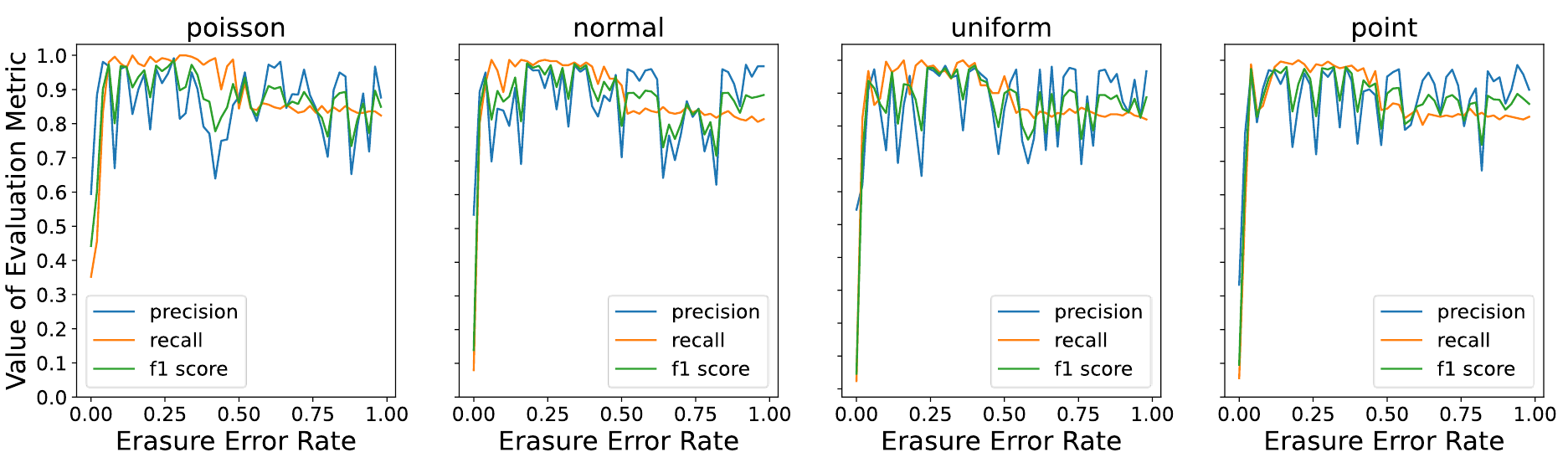}
  \caption{Sensitivity and accuracy of the Autoencoder with erasure errors (for the fourth sensor)}
  \label{fig:autoencoder-erasure-sens}
  \end{figure}

\begin{figure}[h]
  \centering
  \includegraphics[scale=0.275]{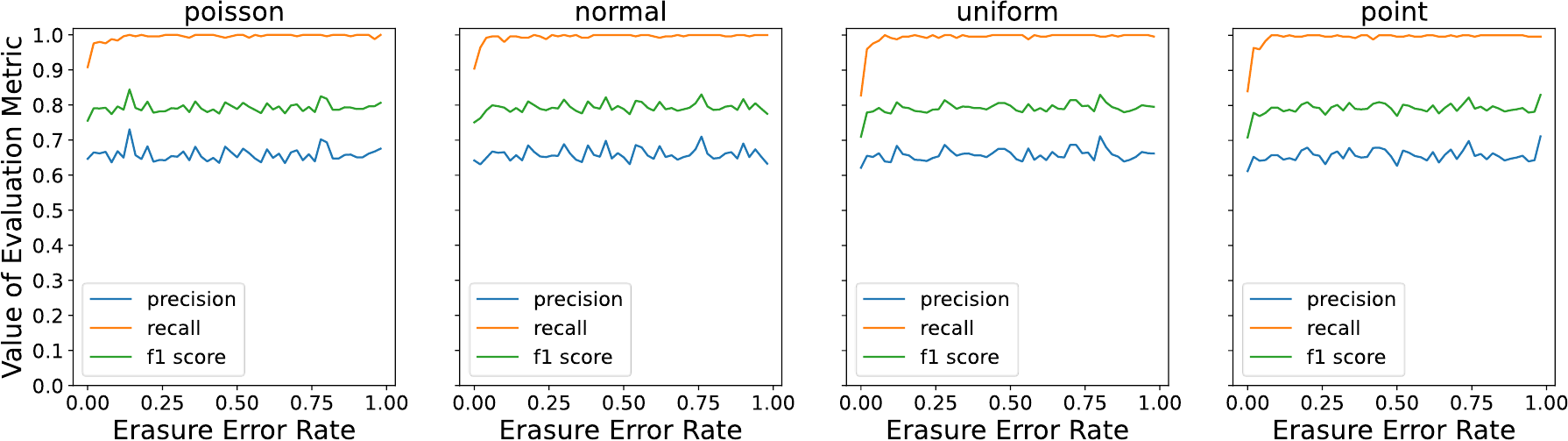}
  \caption{Sensitivity and accuracy of the PCA with erasure errors (for the fourth sensor)}
  \label{fig:pca-erasure-sens}
  \end{figure}

The Poisson distribution presents a unique case as it is a skewed version of the normal distribution. Depending on the skewness, all readings may be classified as anomalous. For this distribution, both PCA and Autoencoder behave similarly, detecting anomalies with noticeable deviations from normal sensor readings. The Autoencoder’s lack of performance drop shows its continued sensitivity to changes in the distribution.

To evaluate sensitivity, the percentage of anomalous readings is varied. The anomalous sensor outputs anomalous readings with probability p and normal readings with probability 1-p, with p ranging from 0 to 1 in intervals of 0.02. This yields 250 + 250*(1-p) normal readings and 250*p anomalous readings. Both PCA and Autoencoder are tested under these conditions.

In Figures \ref{fig:autoencoder-erasure-sens} and \ref{fig:pca-erasure-sens}, the x-axis represents p (probability of anomalous readings), and the y-axis represents precision, recall, and F1 score. A dip in these metrics indicates the algorithm’s inability to detect anomalies. The PCA detects anomalies with high precision and recall even at low values of p, while the Autoencoder is less sensitive, identifying anomalies only as p increases.

Zooming in on p between 0 and 0.1 with intervals of 0.002 shows that PCA detects anomalies when only 0.5\% of readings are anomalous (p = 0.005), while Autoencoder requires around 2.5\% anomalous readings (p = 0.025). This confirms that PCA is more sensitive than Autoencoder. However, PCA’s higher sensitivity could lead to false positives, especially with low anomaly percentages, where the Autoencoder is more robust. This highlights the value of the hybrid approach, which combines the strengths of both models to detect anomalies while minimizing false positives.

\subsection{Evaluation of the Proposed Hybrid Model}\label{sec:results3}

In order to best evaluate the proposed hybrid model, we look at two of the most important metrics: effectiveness as measured by F1 score, and average runtime.

\subsubsection{Comparing Anomaly Detection Effectiveness of Hybrid Model}\label{sec:results3a}

As can be seen in Figure \ref{fig:runtime-table}, the UMAP was by far the slowest algorithm despite it being much faster than TSNE. The Autoencoder performed the best in terms of F1 score but was 2 orders of magnitude slower than the PCA, which had the worst F1 score of all the approaches. The effectiveness of our Hybrid model is clear, as it has a similar F1 score to the Autoencoder but performs 35\% faster.

\begin{table}[!t]
\caption{Runtime Comparision}
\label{fig:runtime-table}
\centering
\begin{tabular}{l c c}
\hline
\textbf{Technique} & \textbf{Runtime} & \textbf{F1-Score} \\
\hline
UMAP         & 64.4 ms  & 0.83666 \\
PCA          & 0.031 ms & 0.63984 \\
Autoencoder  & 2.84 ms  & 0.85566 \\
Hybrid       & 1.84 ms  & 0.84892 \\
\hline
\end{tabular}
\end{table}

\subsubsection{Runtime Analysis of the Hybrid Model}\label{sec:results3b}

\begin{figure}[h]
  \centering
  \includegraphics[scale=0.3]{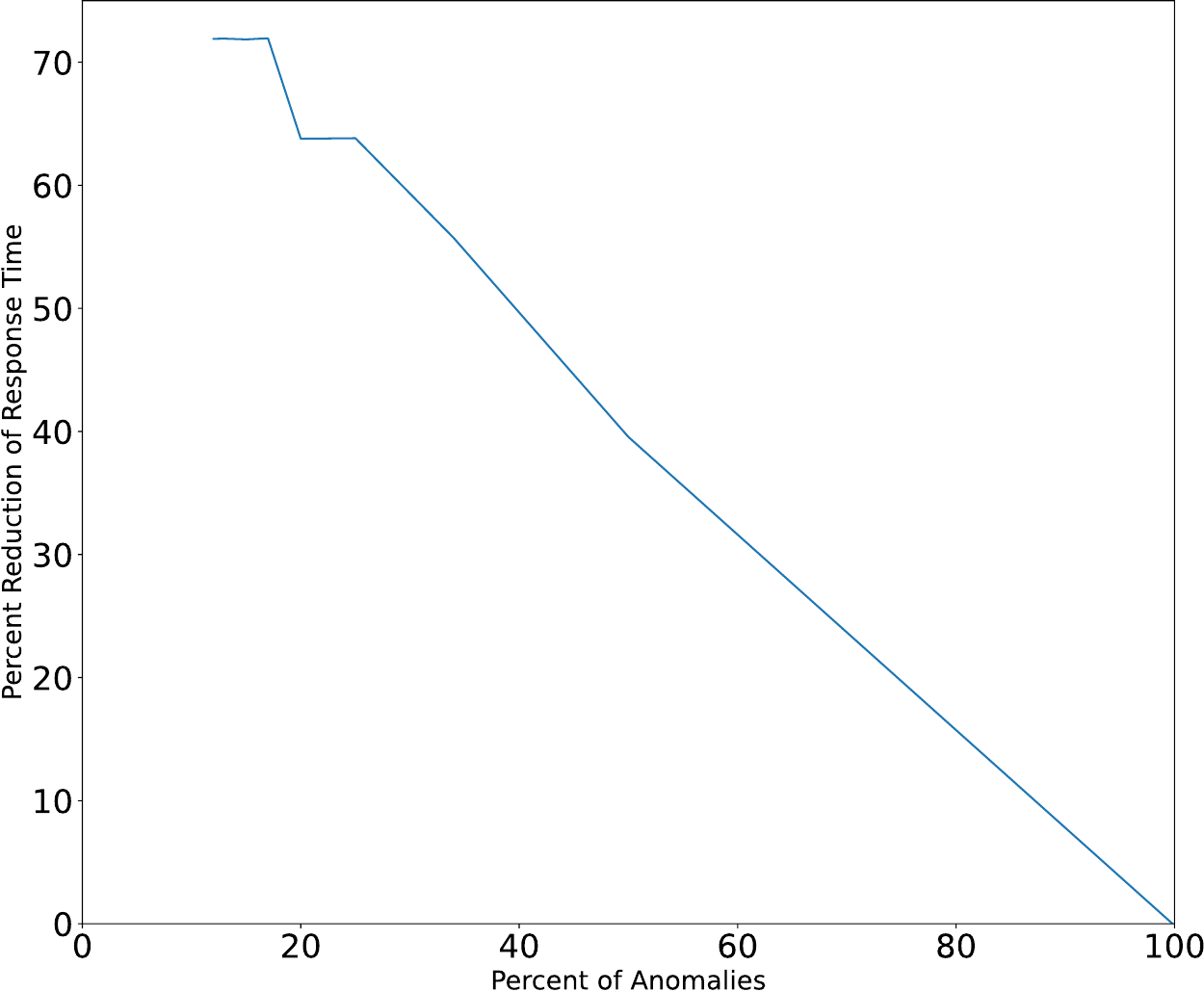}
  \caption{Percent reduction of the response time of the hybrid model versus the Autoencoder model}
  \label{fig:percent-reduction}
  \end{figure}

As shown in Figure \ref{fig:percent-reduction}, there is a strong negative correlation (R = -0.9971) between response time and the percentage of anomalies, meaning the Hybrid approach is faster when anomalies are rare, making it highly efficient in practice.

From Table \ref{fig:runtime-table} and Figure \ref{fig:percent-reduction}, it is obvious that the response time of the hybrid approach is significantly better than the Autoencoder while taking advantage of Autoencoder’s immunity to false positives. This hybrid approach is also resource efficient as both the PCA and the Autoencoder reduce the resource burden associated with high dimensionality.

\section{Conclusion}\label{sec:conclusion}
This research introduces a novel hybrid machine learning approach for anomaly detection in correlated sensor data streams, combining PCA and Autoencoder. PCA is used for fast, continuous monitoring of sensor readings, while Autoencoder is selectively employed for deeper anomaly detection when significant deviations occur. Extensive simulations with real-world Intel Berkeley Research Labs data and synthesized datasets demonstrate that the hybrid method effectively balances computational overhead, response time, and accuracy.

The results show that PCA is highly sensitive to small variations, enabling rapid anomaly detection, while Autoencoder improves robustness against minor fluctuations and reduces false positives. This hybrid model achieves an F1 score comparable to the standalone Autoencoder, with significantly reduced runtime. Compared to existing methods like UMAP, PCA, and Autoencoder alone, the hybrid approach offers superior precision and scalability, making it well-suited for resource-constrained IoT environments.

\end{document}